\documentclass[10pt]{iopart}

\usepackage{hyperref}
\hypersetup{colorlinks=true,       
urlcolor=blue,          
linkcolor=blue,         
citecolor=blue,         
filecolor=blue,          
urlcolor=blue}

\usepackage{graphicx}

\usepackage{harvard}
\citationmode{abbr}
\usepackage{iopams}  

\usepackage{xr}
\usepackage{lineno}

\begin{document}

\ \\\textbf{ACCEPTED MANUSCRIPT}\\ \ \\
This Accepted Manuscript is available for reuse under a CC BY-NC-ND 3.0 licence after the 12 month embargo period provided that all the terms of the licence are adhered to.\\  \ \\
Published in final edited form as:\\
Chella, F., Pizzella, V., Zappasodi, F., Marzetti, L. (2016). Impact of the reference choice on scalp EEG connectivity estimation. \textit{Journal of Neural Engineering}, 13, 036016. DOI: \href{https://doi.org/10.1088/1741-2560/13/3/036016}{https://doi.org/10.1088/1741-2560/13/3/036016}\\ \ \\ 
Copyright information and permissions:\\
\href{https://publishingsupport.iopscience.iop.org/copyright-journals/}{https://publishingsupport.iopscience.iop.org/copyright-journals/} 
\vspace{0cm}

\title[]{Impact of the reference choice on scalp EEG connectivity estimation}

\author{Federico Chella$^{1,2}$, Vittorio Pizzella$^{1,2}$, Filippo Zappasodi$^{1,2}$ and Laura Marzetti$^{1,2}$}
\address{$^1$ Department of Neuroscience, Imaging and Clinical Sciences, ``G. d'Annunzio" University of Chieti-Pescara, Chieti, Italy}
\address{$^2$ Institute for Advanced Biomedical Technologies, ``G. d'Annunzio" University of Chieti-Pescara, Chieti, Italy}
\ead{f.chella@unich.it}

\begin{abstract}

\noindent\textit{Objective}. Several scalp EEG functional connectivity studies, mostly clinical, seem to overlook the reference electrode impact. The subsequent interpretation of brain connectivity is thus often biased by the choice a non-neutral reference. This study aims at systematically investigating these effects.\\
\textit{Approach}. As EEG reference, we examined: the vertex electrode (Cz); the digitally linked mastoids (DLM); the average reference (AVE); and the Reference Electrode Standardization Technique (REST). As a connectivity metric, we used the imaginary part of coherency. We tested simulated and real data (eyes open resting state), by evaluating the influence of electrode density, effect of head model accuracy in the REST transformation, and impact on the characterization of the topology of functional networks from graph analysis.\\
\textit{Main results}. Simulations demonstrated that REST significantly reduced the distortion of connectivity patterns when compared to AVE, Cz and DLM references. Moreover, the availability of high-density EEG systems and an accurate knowledge of the head model are crucial elements to improve REST performance, with the individual realistic head model being preferable to the standard realistic head model. For real data, a systematic change of the spatial pattern of functional connectivity depending on the chosen reference was also observed. The distortion of connectivity patterns was larger for the Cz reference, and progressively decreases when using the DLM, the AVE, the REST. Strikingly, we also showed that network attributes derived from graph analysis, i.e.~node degree and local efficiency, are significantly influenced by the EEG reference choice.\\
\textit{Significance}. Overall, this study highlights that significant differences arise in scalp EEG functional connectivity and graph network properties, in dependence of the chosen reference. We hope our study will convey the message that caution should be taken when interpreting and comparing results obtained from different laboratories when using different reference schemes.\\

\end{abstract}


\maketitle



\section{Introduction}
\label{sec_intro}

The organization of neuronal communication, integration, and functional binding in the brain is one of the central questions of neuroscience. Indeed, in the last decade it has become clear that an adequate picture of brain functioning can be obtained only by understanding the brain as a complex structural and functionally integrated system. Despite this concept is well defined, the idea of brain connectivity in neuroscience refers to several different and interrelated aspects of brain organization \cite{horwitz03,friston11} that are well suited to be investigated with various structural or functional neuroimaging modalities. Electroencephalography (EEG), with its excellent temporal resolution, is a valuable and cost-effective tool for the study of brain functional interactions in a wide range of clinical and research applications \cite{friston95,courchesne05,stam07b,fogelson13,frantzidis14,vanschependom14} since it offers a window into the spatiotemporal structure of phase-coupled cortical oscillations which have been hypothesized to serve as a mechanism for neuronal communication \cite{tallonbaudry96,gross06,womelsdorf06,fries09,miller09}. The recent advances in EEG recording technologies, such as the development of high-density EEG systems, have allowed for increased topographic accuracy, with improved data quality and reduced preparation time \cite{tucker93,holmes10,kleffnercanucci12}. Additionally, the opportunity to combine scalp EEG with other imaging modalities, as well as with robotics or neurostimulation, has made this technique more attractive for many emerging research fields \cite{lebedev06,wolpaw11,bestmann13}.

Despite of the enormous technological advances, however, an old technical issue, namely the choice of the EEG reference, still lacks an accepted solution. This issue arises from the fact that, since only relative measures of electric potential are possible, the EEG signals represent the potential difference between each location over the scalp where the EEG electrodes are placed and a reference site. The latter should be an electrically neutral location to avoid any contamination of the signal of interest by the reference activity. However, there are not neutral locations in the human body \cite{nunez06}, and any choice for the reference location inevitably affects the EEG measurements. In order to minimize this effect, a number of different reference schemes have been proposed, including the vertex \cite{lehmann98,hesse04}, nose \cite{andrew96,essl98}, neck ring \cite{katznelson81}, uni-mastoid or ear \cite{basar98,thatcher01}, linked mastoids or ears \cite{gevins00,croft02}, average reference (i.e.~average potential over all EEG electrodes) \cite{offner50,nunez01}, which provide a “relatively” neutral reference, at least with respect to the signal of interest. The issue of which of the above references is least biased and thus most appropriate for EEG measurement has long been debated \cite{kayser10,nunez10} with the preferential use of one referencing scheme over the others leading to de facto conventions for specific laboratories, research fields or clinical practices. The lack of an universally accepted reference scheme also represents a major obstacle for across-study comparability \cite{kayser10}. In this framework, the average reference has obtained large consensus thanks to a number of objective advantages over the other referencing strategies \cite{srinivasan98,ferree06,nunez06}. The main reason comes from the observation that the surface integral of the electric potential over a volume conductor containing all the current sources is zero \cite{bertrand85}. Thus, the average potential over all the electrodes provides a virtual zero-potential point, insofar as it approximates this integral. An alternative approach was later proposed by \citeasnoun{yao01} with the Reference Electrode Standardization Technique (REST). REST transforms the EEG potentials referenced to any scalp point (or to a combination of them, such as the average) into the potentials referenced to a point located at infinity, far from all the possible neuronal sources and thus acting as an ideal neutral reference location. Despite the proven advantages \cite{srinivasan98,yao01,ferree06,nunez06,marzetti07,yao07,qin10}, however, the latter two approaches are also not completely free of limitations, mainly due to an insufficient electrode density, scalp coverage or, additionally and solely for the REST, to an inaccurate knowledge of the head model \cite{desmedt90,dien98,junghofer99,yao01,zhai04,liu15}.

Given the above considerations, our major concern at this point is not the search for the ideal neutral reference, rather the possible consequence in the analysis and interpretation of EEG data and functional connectivity induced by the chosen reference scheme. Indeed, the reference choice affects both spatial and temporal features of the recorded scalp potentials. In relation to the former, the effects of the reference on the shape of EEG potential maps turn into the sum or subtraction of a constant value to all the electrodes. This was nicely depicted as the effect of rising or receding the water level of a lake in a mountainous area, which changes the location of the zero water level mark, but not the landscape \cite{geselowitz98}. The effects on the temporal aspects of the EEG data are even more marked, due to the fact that a non-neutral reference introduces some time-varying activity to the recordings at all the electrodes. This not only induces a distortion of the temporal waveforms of the EEG recordings, but also an alteration of their spectral properties, e.g.~power spectrum, which is often less intuitive due to the required transformation. When it comes to estimating functional connectivity from electric scalp potentials, the addition of some activity to all the electrodes has the ultimate severe effect of creating spurious connections or suppressing existing ones. To date, relatively few studies have systematically investigated this effect. For instance, it has been shown that EEG correlation \cite{rummel07,muller14} or coherence \cite{fein88,andrew96,essl98} are artificially inflated or deflated by the reference activity contributing to both of the signals involved in the estimation. Analogous results have been reported for the estimation of phase coherency \cite{guevara05,schiff05}. \citeasnoun{zaveri00} investigated the effects on functional connectivity estimated from invasive intracranial EEG referenced to a scalp electrode signal, such as the one recorded from a single mastoid, and reported an increase in the magnitude squared coherence due to the contamination of the reference signal by artifactual activity. \citeasnoun{marzetti07} and \citeasnoun{qin10} reported bias effects on functional connectivity measured as coherence or imaginary coherency for various references, including the REST transformation for a spherical head model. More recently, \citeasnoun{cohen14} found a striking difference among various spatial transformations (reference-based and reference-free) in connectivity analyses through inter site phase clustering \cite{gulbinaite14}. To the specific aim of studying EEG phase synchrony in infants, \citeasnoun{tokariev15} highlighted a dependency on the analysis montage for phase synchrony through imaginary phase locking value \cite{vinck11}. Taken together, these studies point towards a clear effect of the reference choice on functional connectivity results, which, in turn, suggests that researchers may come to different conclusions when interpreting connectivity results obtained from different reference schemes. Along this line, it is also conceivable that the observed effects will impact network properties derived from graph theoretical analysis. Although graph theory is a widely used tool to assess functional networks, the influence of the reference choice on network properties has, to our knowledge, only been reported in \citeasnoun{qin10}, where the authors found changes in network pattern and weighted density depending on the used reference.

However, in spite of the fact that the choice of the EEG reference has been proven to have significant effects on the estimation of functional connections, it was and still is common to find studies in which functional connectivity is estimated from scalp EEG without using the REST or at least the average reference, a choice which poses the question of possible distortions in the connectivity estimates due to the referencing scheme \citeaffixed{shinosaki03,sauseng05,leistedt09,hori13,cavinato15,li15,alba16,ligeza16,naro16}{e.g.,}.

The aim of this paper is to contribute in the above direction by a quantitative investigation of the effects of the reference choice on EEG functional connectivity estimation, through simulated and real data. To this end, we considered the vertex (Cz), the digitally linked mastoid, the average reference, and the REST transformation as possible reference schemes, and the imaginary coherency \cite{nolte04,marzetti07,marzetti08,nolte10} as a connectivity metric. In simulations, we also investigated the effects of the reference choice on EEG potentials for a direct comparison with previous studies \cite{yao01,zhai04,marzetti07,liu15}. By using simulations, we evaluated the influence of the electrode density in the performance of the above references. In addition, since it has been highlighted the need of using a realistic head model to improve the performances of REST in re-referencing potentials \cite{liu15}, we investigated the effects of possible inaccuracies in realistic head model construction, which arise due to, e.g., finite resolution of head structural images, or when a standard realistic head model is used in place of the actual head model, as common practice. The contrast with the ideal three-shell spherical head model \cite{yao01,yao05} was also included for comparison with previous results. Finally, we used real data to evaluate the impact of the references on the characterization of the topology of functional networks from graph analysis applied to the EEG.


\section{Material and methods}
\label{sec_method}

\subsection{The EEG references}

This section gives an overview of the reference schemes most commonly used in EEG studies. Notation and formal definitions are also introduced for later use in this paper.

\subsubsection{Cephalic electrode reference to Cz.}

The reference to a common cephalic electrode is probably the simplest choice for the EEG reference electrode. In such an arrangement, all the electrodes measure the electric potential difference between the electrode site and the reference site. Since any location over the scalp is far from being electrically neutral, it is well recognized that the activity at the reference site is contributing to all the recordings.

In this work, the vertex electrode (Cz) (see figure \ref{electrode_arrays}) is used as cephalic reference electrode. This choice is commonly adopted as on-line reference. In any case, if another reference is chosen during data acquisition, data can always be re-referenced to Cz through an off-line transformation. Specifically, if we denote by $\mathbf{V}_{m}$ the $N\times M$ matrix whose rows contain the EEG recordings measured with any original reference, i.e.~with $N$ being the number of electrodes and $M$ being the number of time samples, then $\mathbf{V}_{m}$ can be re-referenced to Cz by subtracting, for each time sample, the potential measured at Cz from each channel. This is equivalent to applying to the original data the following linear transformation:
\begin{equation}
\mathbf{V}_{Cz} = \mathbf{T}_{Cz}\mathbf{V}_{m} = (\mathbb{I}-\mathbf{R}_{Cz})\mathbf{V}_{m}
\label{Cztrafo}
\end{equation}
where $\mathbb{I}$ is the $N\times N$ identity matrix and
\begin{equation}
\mathbf{R}_{Cz} = \left[ 
\begin{array}{c c c c c c c c c}
0 & 0 & \dots & 1 & \dots & 0 & 0 \\
0 & 0 & \dots & 1 & \dots & 0 & 0 \\
\vdots & \vdots & \vdots & \vdots & \vdots & \vdots & \vdots \\
0 & 0 & \dots & 1 & \dots & 0 & 0 
\end{array}\right]
\end{equation}
is a $N\times N$ matrix with non-zero entries, i.e.~1, only at the column corresponding to the Cz electrode.

\subsubsection{Digitally linked mastoids reference.}
The digitally linked mastoids (DLM) reference is another popular choice for the reference. It consists in a virtual reference obtained by averaging the potentials recorded at the left and right mastoids. Similarly to the Cz reference, the DLM reference can be obtained from any original reference by subtraction. Specifically, for each time sample, half of the potential difference between the electrodes located at the left and right mastoids is subtracted from each channel. The corresponding linear transformation is:
\begin{equation}
\mathbf{V}_{DLM} = \mathbf{T}_{DLM}\mathbf{V}_{m} = (\mathbb{I}-\mathbf{R}_{DLM})\mathbf{V}_{m}
\label{DLMtrafo}
\end{equation}
where $\mathbb{I}$ is the $N\times N$ identity matrix and
\begin{equation}
\mathbf{R}_{DLM} = \left[ 
\begin{array}{c c c c c c c c c}
0 & 0 & \dots & 0.5 & \dots & 0.5 & \dots & 0 & 0 \\
0 & 0 & \dots & 0.5 & \dots & 0.5 & \dots & 0 & 0 \\
\vdots & \vdots & \vdots & \vdots & \vdots & \vdots & \vdots & \vdots & \vdots \\
0 & 0 & \dots & 0.5 & \dots & 0.5 & \dots & 0 & 0
\end{array}\right]
\end{equation}
is a $N\times N$ matrix with non-zero entries, i.e.~0.5, only at the columns corresponding to the electrodes located in the proximity of the left and right mastoids. 

\subsubsection{Average reference.}
The average (AVE) reference, also called common average reference (CAR), consists in referencing the EEG potentials to the average potential of all the electrodes. The AVE reference can be computed by subtracting, for each time sample, the average of all the electrodes from each channel. The corresponding linear transformation is:
\begin{equation}
\mathbf{V}_{AVE} = \mathbf{T}_{AVE}\mathbf{V}_{m} = (\mathbb{I}-\mathbf{R}_{AVE})\mathbf{V}_{m}
\label{AVEtrafo}
\end{equation}
where $\mathbb{I}$ is the $N\times N$ identity matrix and
\begin{equation}
\mathbf{R}_{AVE} = \left[ 
\begin{array}{c c c c c c c c c}
1/N & 1/N  & \dots  & 1/N  & 1/N  \\
1/N & 1/N  & \dots  & 1/N  & 1/N  \\
\vdots & \vdots & \vdots & \vdots & \vdots  \\
1/N & 1/N  & \dots  & 1/N  & 1/N 
\end{array}\right]
\end{equation}
is a $N\times N$ matrix with all the entries equal to $1/N$.

\subsubsection{Reference Electrode Standardization Technique.}
\label{sec:REST}
The Reference Electrode Standardization Technique (REST) \cite{yao01} aims at constructing a virtual reference to a point located at infinity. REST exploits the fact that EEG potentials measured with any original reference and those referenced to a point at infinity are generated by the same (unknown) neuronal sources. Then, if we denote by $\mathbf{S}$ the unknown matrix of the source activities and by $\mathbf{G}_{REST}$ the transfer matrix from these sources to sensors with a reference point at infinity, we have
\begin{equation}
\mathbf{V}_{REST} = \mathbf{G}_{REST}\mathbf{S}
\label{Vrest}
\end{equation}
where $\mathbf{V}_{REST}$ denotes the matrix of the reconstructed EEG recordings referenced to a point at infinity. A similar expression holds for the EEG recordings measured with any original reference, i.e.
\begin{equation}
\mathbf{V}_{m} = \mathbf{G}_{m}\mathbf{S}
\label{Vx}
\end{equation}
where $\mathbf{G}_{m}$ is the corresponding transfer matrix. Thus, by combining the above equations, it is possible to derive a linear transformation $\mathbf{T}_{REST}$  that allows to directly estimate $\mathbf{V}_{REST}$ from $\mathbf{V}_{m}$ as in:
\begin{equation}
\mathbf{V}_{REST} = \mathbf{G}_{REST}\mathbf{S} = \mathbf{G}_{REST}(\mathbf{G}_{m}^{+}\mathbf{V}_{m}) = \mathbf{T}_{REST}\mathbf{V}_{m}
\label{RESTtrafo}
\end{equation}
where $(\cdot)^+$ denotes the Moore-Penrose generalized inverse and
\begin{equation}
\mathbf{T}_{REST}=\mathbf{G}_{REST}\mathbf{G}^+_{m}
\label{Trest}
\end{equation}

The main advantage of REST is that we do not need to explicitly solve the EEG inverse problem, that is, we do not need to know the actual sources $\mathbf{S}$ to compute the transformation matrix $\mathbf{T}_{REST}$. Indeed, from equation (\ref{Trest}), we observe that only the transfer matrices $\mathbf{G}_{REST}$ and $\mathbf{G}_{m}$ are needed to build $\mathbf{T}_{REST}$. Since the potential generated by any source can be equivalently produced by a source distribution enclosing the actual sources \cite{yao00,yao03}, we may assume, for instance, an equivalent source distribution (ESD) on the cortical surface which encloses all the possible neural sources, and calculate $\mathbf{G}_{REST}$ and $\mathbf{G}_{m}$ based on this ESD rather than on the actual sources. The major advantage of this approach is that $\mathbf{T}_{REST}$ does not depend on actual EEG data, but only on the head model, electrode montage, original reference and the spatial geometric information of the assumed ESD. In this study the ESD was assumed to be a discrete layer of current dipoles forming a closed surface, in analogy with previous studies \citeaffixed{marzetti07,yao01,yao05}{e.g.}.

\subsection{Connectivity estimation by imaginary part of coherency}
In the present study, the imaginary part of coherency \cite{nolte04} is used as a measure of functional connectivity between EEG signals. We here briefly recall its definition and properties for later use in this paper.

Let $v_i(t)$ and $v_j(t)$ be the time series of signals recorded by two EEG electrodes, namely $i$ and $j$,  with any given reference. Their cross-spectrum is defined as
\begin{equation}
C_{ij}(f) =\left<\hat{v}_i(f) \hat{v}_j^*(f)\right>
\label{eq:cs0}
\end{equation}
where $\hat{v}_i(f)$ and $\hat{v}_j(f)$ are the complex-valued Fourier coefficients of (eventually windowed) data segments in which $v_i(t)$ and $v_j(t)$ are divided, $^*$ denotes complex conjugation, and $\left<\cdot\right>$ denotes expectation value, i.e.~the average over a sufficiently large number of segments. Coherency is defined as the cross-spectrum normalized by power, i.e.
\begin{equation}
Coh_{ij}(f) =\frac{C_{ij}(f)}{\left(C_{ii}(f)C_{jj}(f)\right)^{1/2}}
\label{eq:cohy1}
\end{equation}
whose magnitude ranges from 0 to 1.

Coherency is a complex-valued function of frequency, and is essentially a measure of how the phases of signals at a specific frequency are coupled. To see this, let us rewrite the Fourier coefficients $\hat{v}_i(f)$ and $\hat{v}_j(f)$ in terms of their amplitude and phases (in the following, the dependence on frequency will be omitted for the ease of reading), i.e.~$\hat{v}_i=a_i\exp(\imath \varphi_i)$ and $\hat{v}_j=a_j\exp(\imath \varphi_j)$, with $\imath$ being the imaginary unit. Then coherency becomes
\begin{equation}
Coh_{ij}(f) =\frac{\left<a_i a_j \exp(\Delta\varphi)\right>}{\left(\left<a_i^2\right>\left<a_j^2\right>\right)^{1/2}}
\end{equation}
where $\Delta \varphi=\varphi_i - \varphi_j$ denotes the phase difference between the signals recorded by electrodes $i$ and $j$ at a specific frequency. It turns out that coherency is the expectation value of $\Delta \varphi$ weighted with the product of the signal amplitudes, apart for a normalization factor. If the phases of the two signals are not coupled, $\Delta \varphi$ is a random number, and thus coherency vanishes.

A serious concern for scalp EEG connectivity analysis is represented by the artifacts of volume conduction \cite{nunez97,schoffelen09,srinivasan07,winter07}. These are essentially due to the low spatial resolution of the EEG, namely two sensors can record from the same brain area, opening the possibility for spurious interactions between sensors even in the absence of true brain interactions. Almost all the measures of EEG connectivity, including EEG coherency, are highly sensitive to mixing artifacts.\\
To address the issue in relation to coherency, it has been suggested to use the imaginary part of coherency, inasmuch this quantity is robust to artifact of volume conduction \cite{nolte04}. Indeed, a nonvanishing imaginary part of coherency requires a non-zero value for $\Delta\varphi$ to be observable. Thus, it cannot be generated by the superposition of independent neuronal sources, regardless of the number of sources and how they are mapped into sensors, provided that this mapping is instantaneous, i.e.~with no phase distortions, which is in fact an excellent approximation for frequencies below 1 KHz \cite{stinstra98}. This obviously implies that the imaginary part of coherency is only sensitive to processes time-lagged to each other, whereas perfectly synchronous sources, i.e.~for which $\Delta\varphi =0$, do not contribute to the imaginary part of coherency but only to its real part and therefore cannot be detected using the imaginary part of coherency alone. Also, by considering only the imaginary part of coherency, which measures amplitude weighted phase coupling, it is not possible to differentiate between a change in the magnitude of coherency (i.e., coherence) and a change in the phase relationship, because the magnitude and the phase of complex-valued coherency require both real and the imaginary part to be reconstructed. Demanding to observe this difference and still retaining the robustness to volume conduction would require the application of non-linear methods \cite{chella14}, which is subject to ongoing research.

\subsection{Simulations}
\label{sec:simul}

A simulation study was carried out to investigate the effects of the reference choice on EEG potentials and functional connectivity. Since a point located at infinity would act as an ideal neutral reference, we quantitatively evaluated these effects by contrasting the potentials and imaginary part of coherency values obtained from the EEG referenced to each of the above described references to the same quantities obtained from the EEG referenced to infinity.

\subsubsection{Head model and EEG electrodes.}
\label{sec:headmodels}
Ten different realistic head shapes were used in our simulation in order to take into account possible effects induced by individual anatomical features, as will be motivated in more detail in section \ref{sec:simul_rep_stat} describing the simulation repetitions.
Specifically, the head shapes were obtained from the segmentation by Curry 6.0 software (Neuroscan Compumedics USA, Ltd. - Charlotte, NC, USA) of the MRI whole-head images of the ten subjects recruited for the real data experiment described in this paper (see section \ref{real_data_acquisition}). For each subject, a realistically shaped head model was constructed, consisting of a volume conductor and a source space. The volume conductor included three compartments, i.e.~brain, skull and scalp, while the source space consisted in a three-dimensional grid uniformly sampled in the volume encompassed by the cortical mantle, with a 5 mm step. Relative conductivities were assumed equal to 1.0 for the brain and the scalp, and 0.02=1/50 for the skull. The spatial resolution for the shells delimiting the above compartments, i.e.~inner skull, outer skull and skin, were set equal to 5, 7 and 8 millimetres, respectively.

\begin{figure*}[!htb]
\centering
\includegraphics[width=13cm]{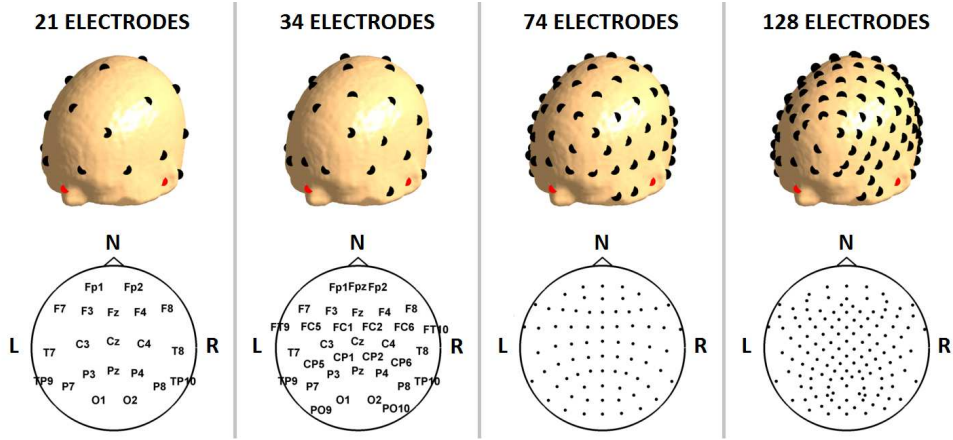}
\caption{\small{EEG electrode layouts used in simulations. From left to right: 21 electrodes, i.e.~19 electrodes located in accordance with the International 10-20 system with the addition of the TP9 and TP10 electrodes; 34 electrodes, i.e.~a selection of the electrodes from the 10-10 system; 74 electrodes, i.e.~the whole 10-10 electrode system; and 128 electrodes, i.e.~a selection of the 10-5 electrode system.}}
\label{electrode_arrays}
\end{figure*}

The full EEG sensor layout consisted of 128 electrodes which were fitted on the outermost shell of the head model, in accordance with the positions provided by the 10-5 electrode system \cite{oostenveld01}. Since the performances of the AVE and the REST are expected to depend on the electrode density \citeaffixed{nunez06,liu15}{e.g.}, different EEG layouts were realized by varying the number of electrodes over the scalp and used in the simulation study. Specifically, we considered four different layouts:
\begin{itemize}
\item 21 electrodes, i.e.~19 electrodes located in accordance with the International 10-20 system \cite{jasper58} with the addition of the TP9 and TP10 electrodes;
\item 34 electrodes, i.e.~a selection of the electrodes from the 10-10 system \cite{chatrian85};
\item 74 electrodes, i.e.~the whole 10-10 electrode system \cite{chatrian85};
\item 128 electrodes, i.e.~a selection of the 10-5 electrode system \cite{oostenveld01}.
\end{itemize}
A schematic representation of the different electrode layouts is given in figure \ref{electrode_arrays}.
All the electrode subsets provide an approximatively uniform coverage of the whole scalp. Moreover, they have been chosen in such a way that the more dense subset includes the more sparse subset (i.e.~the 128-electrode subset includes the 74- , the 34- and the 21-electrode subset, the 74-electrode subset includes the 34- and the 21-electrode subset, and so on). In this way, the observed differences are ascribable only to different electrode densities, and not to a different coverage of the scalp.

\subsubsection{Generation of simulated EEG recordings.}
\label{sec:gendata}
Given the head model for each of the ten subjects and the locations of the EEG electrodes for one of the above defined layouts over the scalp, 5 minute EEG recordings, sampled at 500 Hz, were simulated by first generating a set of brain sources and then by solving the EEG forward problem. All sources were modeled as single current dipoles randomly located and oriented in the brain volume. The set of sources included 2 interacting sources plus 4 uncorrelated sources, as described below: 
\begin{itemize}
\item \textit{interacting sources}: 2 interacting sources of stochastic activity around 10 Hz. Specifically, we first generated the timecourse of a source, say $s_1(t)$, by band-pass filtering white Gaussian noise around 10 Hz, with 0.5 Hz bandwidth, and then we set the activity of a second source, say $s_2(t)$, to a time-delayed copy of $s_1(t)$, i.e., $s_2(t) = s_1(t-\tau)$. The time delay $\tau$ was 10 milliseconds. For data filtering we used a Butterworth filter, performing filtering in both the forward and reverse directions to ensure zero phase distortion.
\item \textit{noisy sources}: 4 uncorrelated sources of broadband white Gaussian noise between 0.5-100 Hz. The signal-to-noise ratio (SNR) at 10 Hz was set equal to 1, with the SNR being calculated as the ratio between the mean variance across channels of the signal generated by interacting sources and the mean variance of signal generated by noisy sources.
\end{itemize}

The EEG forward problem was solved by using an analytic expansion of the EEG lead field for realistic volume conductors \cite{nolte05}. This approach, for known sources, allows calculating, in an approximate form, the theoretical EEG potential referenced to a point at infinity. In order to reproduce realistic experimental conditions, the EEG recordings were contaminated with a low level of either white Gaussian noise or iso-spectral noise to mimic the instrumentation noise. In particular, the iso-spectral noise was generated from the noiseless EEG recordings by using the method of \citeasnoun{prichard94} for surrogate time-series generation. The SNR of both white Gaussian noise and iso-spectral noise was set to 10. The EEG recordings generated with this procedure are thus the theoretical EEG potentials referenced to a point at infinity that will be used as gold standard for comparison with the other references. In the following, the theoretical reference to a point at infinity will be denoted as INF, and the corresponding potentials as $\mathbf{V}_{INF}$, which actually reads as a $N\times M$ matrix, with $N$ being the number of EEG electrodes and $M$ the number of time samples. 

\subsubsection{Re-referencing the simulated data.}
\label{re-referecing_sec_simul}

Datasets for the different EEG references discussed above were derived for each subset of electrodes. Specifically, the datasets referenced to Cz, DLM and AVE were directly obtained from the recordings referenced to INF, $\mathbf{V}_{INF}$, i.e.~by subtracting from all the electrodes, respectively, the signal at Cz electrode, the average between the signals at TP9 and TP10 electrodes (the latters being located in the proximity of the left and right mastoids), and the average signal over all the electrodes.\\
The REST re-referencing was performed on the datasets previously referenced to Cz, the latter being, among the references concerned in this study, the one which is typically used as online reference for actual EEG measurements. It is known that the performance of REST depends on the accuracy of the head model used for the computation of the transformation matrix, with a more accurate head model resulting in a better potential reconstruction \cite{nunez10,yao01,yao05,zhai04}. In order to investigate this effect, we performed the REST re-referencing using four different head models which deviate, to different extents, from the head model used for the generation of the simulated data, as explained in the following:
\begin{itemize}
\item \textit{Spherical head model}. The first case aims at investigating the condition where no knowledge of subject's head anatomy is available (e.g., no MRI images were acquired) and, thus, a spherical head model is used for the computation of the REST transformation matrix. Specifically, we assumed a volume conductor consisting of three concentric spheres delimiting the brain, the skull and the scalp, with relative conductivities equal to 1.0 (for brain and scalp) and 0.02=1/50 (for skull), while the ESD was constrained over a closed surface formed by a spherical cap and a transverse plane. The dimensions of the three concentric spheres and of the spherical cap were based on standard head dimensions provided by the MNI-152 template \cite{fonov09,fonov11}. The particular choice of a three-concentric-sphere model has been performed in analogy with previous studies investigating the effectiveness of the REST \cite{marzetti07,yao01,yao05}. This case will be referred to as REST \textit{spherical}.

\item \textit{Standard head model}. This case is similar to the previous one, except for a standard realistically shaped head model used in place of the spherical model. Specifically, we used the head model obtained from the segmentation of the MNI-152 template \cite{fonov09,fonov11}. Head tissue relative conductivities were set equal to 1.0 for brain and scalp, and 0.02=1/50 for skull. This case will be referred to as REST \textit{standard}.

\item \textit{Inaccurate head model}. Here we assume that subject's whole-head MRI images are available, and thus an individual (i.e.~per subject) head model can be used for the computation of the REST transformation matrix, but we hypothesize that such a head model is corrupted by possible inaccuracies, e.g.~due to finite MRI spatial resolution or errors in the segmentation of MRI images. This condition was simulated by slightly perturbing the geometry of the head model used for the generation of EEG simulated recordings. Specifically, each vertex of the meshes representing the cortex and the three-shell  volume conductor were shifted by a fixed displacement, i.e.~3 millimetres, in random directions. Tissue relative conductivities were kept equal to 1.0 for brain and scalp, and 0.02=1/50 for skull. This case will be referred to as REST \textit{real perturbed}.

\item \textit{Exact head model}. The last case aims at investigating the condition in which an exact knowledge of the subject's head model is available. This was achieved by considering, in the computation of the REST transformation matrix, the same head model used for the generation of simulated EEG recordings. This case will be referred to as REST \textit{real exact}.
\end{itemize}
For each head model, the REST transformation matrix was computed by assuming an equivalent source distribution (ESD) consisting of 4000 current dipoles randomly located and normally oriented over the cortical surface. Specifically, the chosen number of current dipoles, i.e.~4000, was the result of a preliminary simulation study investigating the effects of the ESD discretization on REST performance, which is described in section \ref{sec:ESDmodel} of the Supplementary Material. The computation of the transformation matrix required roughly 27 seconds on a desktop PC (Intel$^{\circledR}$ i5 - 2400 CPU @ 3.10 GHz; RAM 8 GB). A schematic representation of the ESD and the volume conductor model for all the above cases is given in figure \ref{headmodels} of the Supplementary Material. 

In summary, for each simulated data, seven different datasets were obtained from the re-referencing of $\mathbf{V}_{INF}$, which were denoted as: $\mathbf{V}_{Cz}$, $\mathbf{V}_{DLM}$, $\mathbf{V}_{AVE}$, $\mathbf{V}_{REST\,spherical}$, $\mathbf{V}_{REST\,standard}$, $\mathbf{V}_{REST\,real\, perturbed}$ and $\mathbf{V}_{REST\,real\,exact}$.

\subsubsection{Coherency analysis.}
The generated EEG recordings were divided into 1 second non-overlapping segments. Within each segment, data were Hanning windowed, Fourier transformed and the imaginary part of coherency at 10 Hz was estimated between each pair of EEG electrodes. The end result is a square $N\times N$ matrix, with $N$ being the number of EEG electrodes, where the entry in the $i$-th row and $j$-th column reads as the value of the imaginary part of coherency between the recordings at electrodes $i$ and $j$. Following the notation introduced for potentials, the imaginary part of coherency matrices resulting from different EEG referencing conditions were denoted as: $\mathbf{ImCoh}_{INF}$, $\mathbf{ImCoh}_{Cz}$, $\mathbf{ImCoh}_{DLM}$, $\mathbf{ImCoh}_{AVE}$, $\mathbf{ImCoh}_{REST\,spherical}$, $\mathbf{ImCoh}_{REST\,standard}$, $\mathbf{ImCoh}_{REST\,real\, perturbed}$ and $\mathbf{ImCoh}_{REST\,real\,exact}$.

\subsubsection{Performance criteria.}
For each EEG electrode subset (i.e.~21, 34, 74 and 128 electrodes) and for each re-referencing condition, the distortion of the EEG potentials induced by the reference choice was measured as the relative error ($RE$) between the re-referenced EEG recordings and the EEG recordings referenced at infinity, according to the following definition:
\begin{equation}
RE_{V_X} = \frac{||\mathbf{V}_{X}-\mathbf{V}_{INF}||_F}{||\mathbf{V}_{INF}||_F}
\label{REpot}
\end{equation}
where $||\cdot||_F$ denotes the matrix Frobenius norm and $X$ reads, in turn, Cz, DLM, AVE, REST \textit{spherical}, REST \textit{standard}, REST \textit{real perturbed} and REST \textit{real exact}. 

Similarly, the effects of different EEG references on connectivity analysis were evaluated by defining a relative error for the imaginary part of coherency matrices as: 
\begin{equation}
RE_{ImCoh_X} = \frac{||\mathbf{ImCoh}_{X}-\mathbf{ImCoh}_{INF}||_F}{||\mathbf{ImCoh}_{INF}||_F}
\label{REimcoh}
\end{equation}
with $||\cdot||_F$ and $X$ as in (\ref{REpot}).

\subsubsection{Simulation repetitions and statistics.}
\label{sec:simul_rep_stat}
In order to take into account multiple source configurations, the simulations were performed by randomizing source locations and orientations. Moreover, the shape of the realistic head model used for the generation of simulated EEG data was varied among the 10 different realistic shapes used for this simulation study (see section \ref{sec:headmodels}). In particular, the reason for varying the head shape relies on the fact that the results obtained for the REST in the event that a standard head model is used for the computation of the transformation matrix (i.e.~the REST \textit{standard} condition) might depend on the mismatch between subject's individual anatomy and the standard MNI-152 template, i.e.~the former being used for the EEG data generation and the latter for the EEG data recovering. One-hundred simulations were performed for each different head shape, for a total amount of 1000 simulation repetitions.

The contrast between the different EEG referencing conditions has been performed by looking at the distributions of the relative errors, i.e.~$RE_{V_X}$ and $RE_{ImCoh_X}$, from all simulation repetitions. Statistical analysis for the contrast between referencing conditions consisted in non-parametric paired sample statistics, i.e Wilcoxon signed-rank test. The statistical significance level was set at $p<0.05$.

\subsection{Application to real EEG recordings}
\label{real_data}
To study the effects of the choice of the reference in actual EEG measurements, we analyzed EEG data recorded during eyes open resting state. Specifically, we evaluated the effects on connectivity patterns as revealed by the imaginary part of coherency. Moreover, we investigated whether network properties based on graph theoretical analysis, which can be calculated from coherency patterns, are influenced by the EEG reference choice. 

\subsubsection{Data acquisition and preprocessing.}
\label{real_data_acquisition}
Ten healthy adult subjects (gender: 2F, 8M; age: 20-29 years) were recruited for the experiment. Written consent and local ethical committee agreement were obtained. Subjects were requested to sit in a quiet and dimly lit room and to fix a cross in front of them. Measurements consisted of 10 min of continuous eyes-open resting state activity. The EEG signals were recorded using a 128-sensor HydroCel GSN net (Electrical Geodesics, Inc. - Eugene, OR, USA) referenced to Cz. The electrode impedance was kept below 100 k$\Omega$. Data was sampled at 1 kHz. The locations of the EEG channels on the scalp and of three fiducial points (nasion, left and right pre-auricular point) were measured by a 3D digitizer (Polhemus, Colchester, VT, USA).\\
High resolution whole-head anatomical images were acquired using a 3-T Philips Achieva MRI scanner (Philips Medical Systems, Best, The Netherlands) via a 3D fast field echo T1-weighted sequence (MP-RAGE; voxel size 1 mm isotropic, TR = 8.1 ms, echo time TE = 3.7 ms; flip angle 8$^\circ$, and SENSE factor 2). The coregistration of EEG electrode locations with the MRI volume was performed by aligning the fiducial points in the two modalities.

A preprocessing step was carried out before proceeding with data analysis. The signals from electrodes located over the face and neck were taken out because contaminated by muscular activity. The number of available channels was thus equal to 110. Raw data were band-pass filtered at 0.5-100 Hz. All recordings were visually inspected and the segments of data containing spike artifacts were removed. An Independent Component Analysis (ICA) was also performed for instrumental and biological artifact removal. Specifically, ICA was performed by using the FastICA algorithm with deflationary orthogonalization and tanh nonlinearity \cite{hyvarinen00}. The extracted independent components were visually inspected and classified as artifactual components or as components of brain origin on the basis of their topographies, power spectral densities and timecourses. The independent components classified as artifactual were rejected. Particular attention was paid to the rejection of artifacts from the eyes, heart and neck muscles. For a between-subjects comparison, the channels which were possibly missing from the set of 110 channels (i.e., taken out because extremely noisy or damaged) were interpolated from clean signals by using spherical spline interpolation functions \cite{perrin89} available in the FieldTrip software package \cite{oostenveld11}. Specifically, the interpolation was necessary for only one channel in three out of the ten subjects and for two channels in two out of the ten subjects. Interpolation ensured that, to study the effects of the choice of the EEG reference on connectivity analysis, the full set of 110 electrodes could be taken into account. Both the contrast of the imaginary part of coherency patterns and the contrast of two graph theoretical measures (degree and local efficiency) were thus based on the full square matrices of size 110$\times$110 containing the values of the imaginary part of coherency between all pairs of electrodes. 

\subsubsection{Re-referencing the EEG recordings.}
\label{re-referecing_sec_real}
The EEG signals were acquired using an electrode montage referenced to Cz. Other EEG referencing, i.e., DLM, AVE and REST, were then obtained from the original referential montage by using the transformations (\ref{Cztrafo}), (\ref{DLMtrafo}), (\ref{AVEtrafo}) and (\ref{RESTtrafo}) as already discussed in section \ref{re-referecing_sec_simul} for the re-referencing of simulated data, with one exception in relation to the REST. Specifically, as still being interested in investigating the effects of the head model on the performance of REST, in actual experimental conditions it is not possible to contrast a \textit{real exact} head model with a \textit{real perturbed} head model, the exact knowledge of head geometry and conductivity being allowed only in idealized simulations. Therefore, only one realistic head model will be considered in the following along with the \textit{spherical} and \textit{standard} models, which is the one obtained from the segmentation of individual MRI, and which is presumably equivalent to the \textit{real perturbed} head model hypothesized in simulations. Such a head model will be simply referred to as \textit{real} (rather than \textit{real perturbed} or \textit{real exact}). To perform the REST re-referencing, an equivalent source distribution was assumed consisting of 4000 current dipoles randomly located and normally oriented over the cortical surface, while the relative conductivities were assumed equal to 1.0 for the brain and the scalp and 0.02=1/50 for the skull.  

\subsubsection{Coherency analysis.}
For each re-referenced dataset, the signals were divided into 1 second non-overlapping segments. Within each segment, data were Hanning windowed, Fourier transformed and the imaginary part of coherency was estimated between each electrode pair. The resulting frequency resolution was 1 Hz. The subsequent steps of the coherency analysis were restricted to the Alpha band (8-12 Hz). In particular, we focused around the individual (per-subject) frequency within the Alpha band where the imaginary part of coherency showed its maximum (e.g.~$\sim10$ Hz). The analysis was performed for each subject and for each re-referenced dataset separately. Group level matrices for the imaginary part of coherency for the different references were obtained by averaging over subjects. In order to quantitatively evaluate the differences between the referencing conditions, a dissimilarity measure defined as one minus the squared Pearson correlation coefficient i.e., $d_r=1-r^2$, between the vector-like data obtained by unfolding the matrices in the different referencing conditions was calculated. Note that $d_r$ ranges from 0 (no dissimilarity) to 1 (complete dissimilarity).

\subsubsection{Computation of network properties.}

The application of graph theory to the analysis of EEG connectivity data has been extensively studied and discussed in a number of publications \citeaffixed{rubinov10,stam07}{e.g.} and the interested reader is addressed to those references. However, for the sake of clarity, some basic principles and definitions of graph analysis will be recalled in the following for unweighted and undirected graphs.

The first step for applying unweighted graph theoretical analysis to connectivity matrices is to convert the connectivity matrix into a binary graph. A binary graph is a mathematical representation of a network, which is essentially reduced to a set of nodes (e.g., the EEG electrodes) and undirected connections between them. More specifically, in a binary graph, the connections between nodes either exist or do not exist, i.e.~they do not have graded values. The connection status between two nodes $i$ and $j$ is thus represented by a binary value, i.e.~$a_{ij}$: if two nodes are connected, $a_{ij}=1$ and the nodes are said to be neighbours, otherwise $a_{ij}=0$. The construction of a binary graph from a connectivity matrix is often performed by thresholding the connectivity matrix such that only a given percentage of all the possible connections are retained. Following this approach, the imaginary part of coherency matrices were converted into the corresponding binary graphs by retaining the $25\%$ of strongest (both positive and negative) connections between electrodes.

Once the connectivity matrices have been converted into the corresponding binary graphs, it is possible to characterize a number of attributes, including the degree and the local efficiency. In the following, we will focus on these two attributes, which have been indicated as of primary interest for the study of local properties of functional networks \cite{rubinov10}. The \textit{degree} of connectivity for a node, say $i$, is defined as the total number of connections to other nodes, i.e.
\begin{equation}
k_i = \sum_{j=1}^{N} a_{ij}
\end{equation}
with $N$ being the number of nodes. The functional interpretation of the degree is fairly simple: the value of the degree reflects the importance of an individual node in the network and, for this reason, it is often denoted as measure of node centrality. Less straightforward is the interpretation of the \textit{local efficiency} \cite{latora01}, which is an attribute of a graph's node based on the concept of efficiency of the communication between nodes. The efficiency of the communication between two nodes, i.e.~$e_{ij}$, is computed from their distance, i.e.~$d_{ij}$, which is defined as the length of the shortest among all (direct or indirect) paths linking the two nodes. For a binary graph, the length of a path can be computed as the number of connections it contains. The efficiency can then be defined as the reciprocal of the shortest path, i.e.~$e_{i,j}=1/d_{ij}$. If there is no path, the path length is infinite and, consistently, the efficiency of communication between nodes vanishes. Given the efficiency between two nodes, the local efficiency for the node $i$,  i.e.~$E_{loc\,,i}$, is defined as the average over the efficiencies calculated between all possible pairs of nodes in the sub-graph $\mathbf{A}_i$, where $\mathbf{A}_i$ is  the sub-graph formed only by the nodes connected to $i$, i.e.~its neighbours, but not $i$ itself. In formula, this is:
\begin{equation}
E_{loc\,,i} = \frac{1}{N'(N'-1)} \sum_{j\neq k}\frac{1}{d_{jk}}
\end{equation}
with the indices $i$ and $k$ running over the $N'$ nodes forming the sub-graph $\mathbf{A}_i$. The local efficiency is a measure of how efficient is the communication between the neighbours of a given node when that node is removed \cite{latora01}.\\
For the actual analysis, the degree and the local efficiency were computed from the imaginary part of coherency matrices for each subject separately and for the different EEG referencing with the aim of evaluating possible effects of the reference electrode choice on graph attributes. 

The computation of the degree and local efficiency was performed by using the Brain Connectivity Toolbox (BCT, http://www.brain-connectivity-toolbox.net/) \cite{rubinov10}. The statistical comparison between the different EEG referencing for each graph attribute was performed with a paired sample $t$-tests. The statistical significance of the associated $t$-values was assessed through a non-parametric permutation test in which 10000 random permutation of reference labels were carried out. The permutation test was performed by using the FieldTrip software package \cite{oostenveld11}.


\section{Results}
\subsection{Simulations}
The effects of the reference on EEG potential and connectivity estimation were quantitatively evaluated by measuring the relative error for potentials and for the imaginary part of coherency according to equations (\ref{REpot}) and (\ref{REimcoh}), respectively. The obtained results are presented in the following sections.

\subsubsection{Potentials.}\label{sec:res_pot}
\begin{figure*}[!htbp]
\centering
\includegraphics[width=16cm]{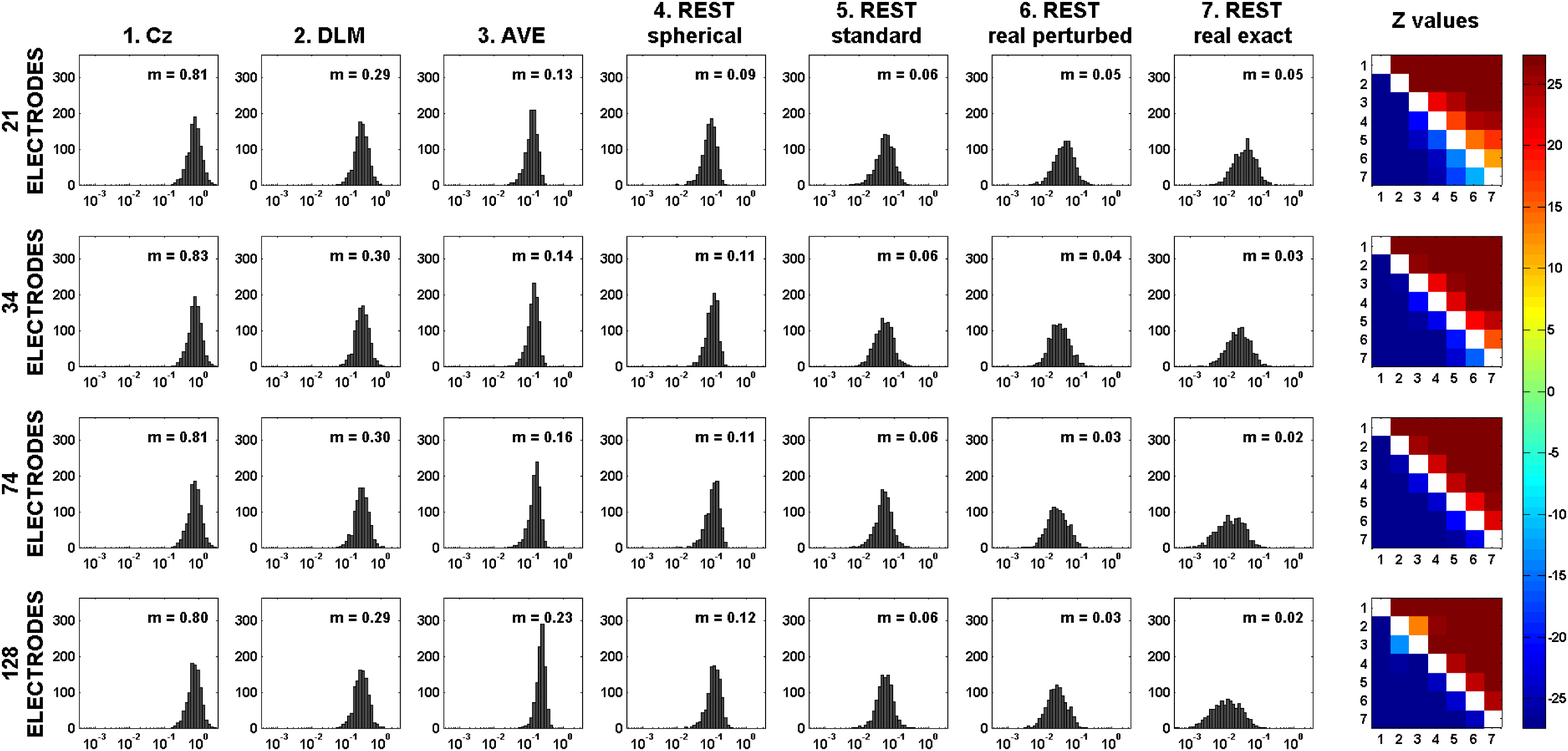}
\caption{\small{Histograms of the relative error ($RE$) for noiseless EEG potentials, i.e.~for all combinations of number of EEG electrodes and EEG referencing conditions. The histograms collect the data from 1000 simulation repetitions. The mean value for $RE$ is denoted by $m$.  For the ease of visualization, the abscissa values for the histograms have been scaled logarithmically. Rightmost side of each panel: z-values for non-parametric paired sample statistics, i.e.~Wilcoxon signed-rank test, performed for the contrast of the $RE$ distributions obtained in different EEG referencing conditions, here labelled with a progressive number from 1 to 7, and for a specific electrode density.}}
\label{noiseless_pot_hist}
\end{figure*}

We first present the results obtained for noiseless simulated potentials. The respective relative errors ($RE$) are shown in figure \ref{noiseless_pot_hist}. In each sub-figure, the histogram collecting the values of $RE$ from the 1000 simulations repetitions is shown for a given combination of electrode number and EEG referencing condition. Specifically, the sub-figures on the same row correspond to the same electrode number, while the sub-figures on the same column correspond to the same referencing scheme. The mean $RE$ value over all the simulation repetitions is indicated in the top-right corner of each sub-figure. In addition, in the rightmost side of figure \ref{noiseless_pot_hist}, the z-values are shown for a non-parametric paired sample statistics, i.e.~Wilcoxon signed-rank test, performed to contrast the $RE$ distributions from the different referencing conditions (here labelled with a progressive number from 1 to 7) and for a specific electrode density.
\\
We observe that, regardless of the number of electrodes, the mean value of $RE$ for the Cz reference ($\sim 0.813$) is much larger than those obtained for the other referencing conditions. Lower values for the mean $RE$ are achieved when using the DLM reference ($\sim 0.295$), although these values are still larger if compared to those obtained for AVE and REST reference. Importantly, the mean value of $RE$ for Cz and DLM reference does not depend on the number of EEG electrodes. The AVE reference turns out to be a better choice than the DLM and Cz reference, as demonstrated by the $RE$ being effectively reduced. We also noticed that the $RE$ values increase from 0.13 (on average) in case of 21 electrodes to 0.23 (on average) in case of 128 electrodes. Finally, the REST largely outperforms all the other references when the head model is accurately known, i.e.~in the REST \textit{real exact} condition. Specifically, $RE\leq0.05$ (on average) and decreases as the number of electrodes increases, i.e.~down to $RE=0.02$ (on average) for the 128 electrode array. If the knowledge of the head model is inaccurate, the performance of the REST progressively worsens, as shown by the increase of the $RE$ obtained for the REST re-referencing using a \textit{real perturbed}, a \textit{standard} or a \textit{spherical} head model. In any case, even when a spherical head model is used, REST performs better than AVE reference, while the benefits of improved electrode density for reducing the $RE$, which were observed in the \textit{real exact} condition, apparently diminish (i.e.~for the \textit{real perturbed} condition) or vanish (i.e.~for the \textit{standard} and \textit{spherical} conditions).\\
Statistical analysis by a Wilcoxon signed-rank test allowed to confirm that the above discussed results are statistically significant as shown in the rightmost side of figure \ref{noiseless_pot_hist}. Here, the z-scores for all the pairwise comparisons of the $RE$ distributions from the different referencing conditions and for a specific electrode density are reported. A positive z-scores indicates that the reference condition listed on the vertical axis exhibits a $RE$ which is larger (in a statistical sense), and thus a worse performance, than the one of the reference condition listed on the horizontal axis, and vice versa for a negative z-score. Clearly, the plotted z-score matrices are antisymmetric, while the different EEG references have been intentionally sorted in ascending performance order. All the z-scores are significant for a p-value $p<0.001$, thus confirming that the effects of using different EEG references are significantly different.

\begin{figure*}[!p]
\centering
\includegraphics[width=16cm]{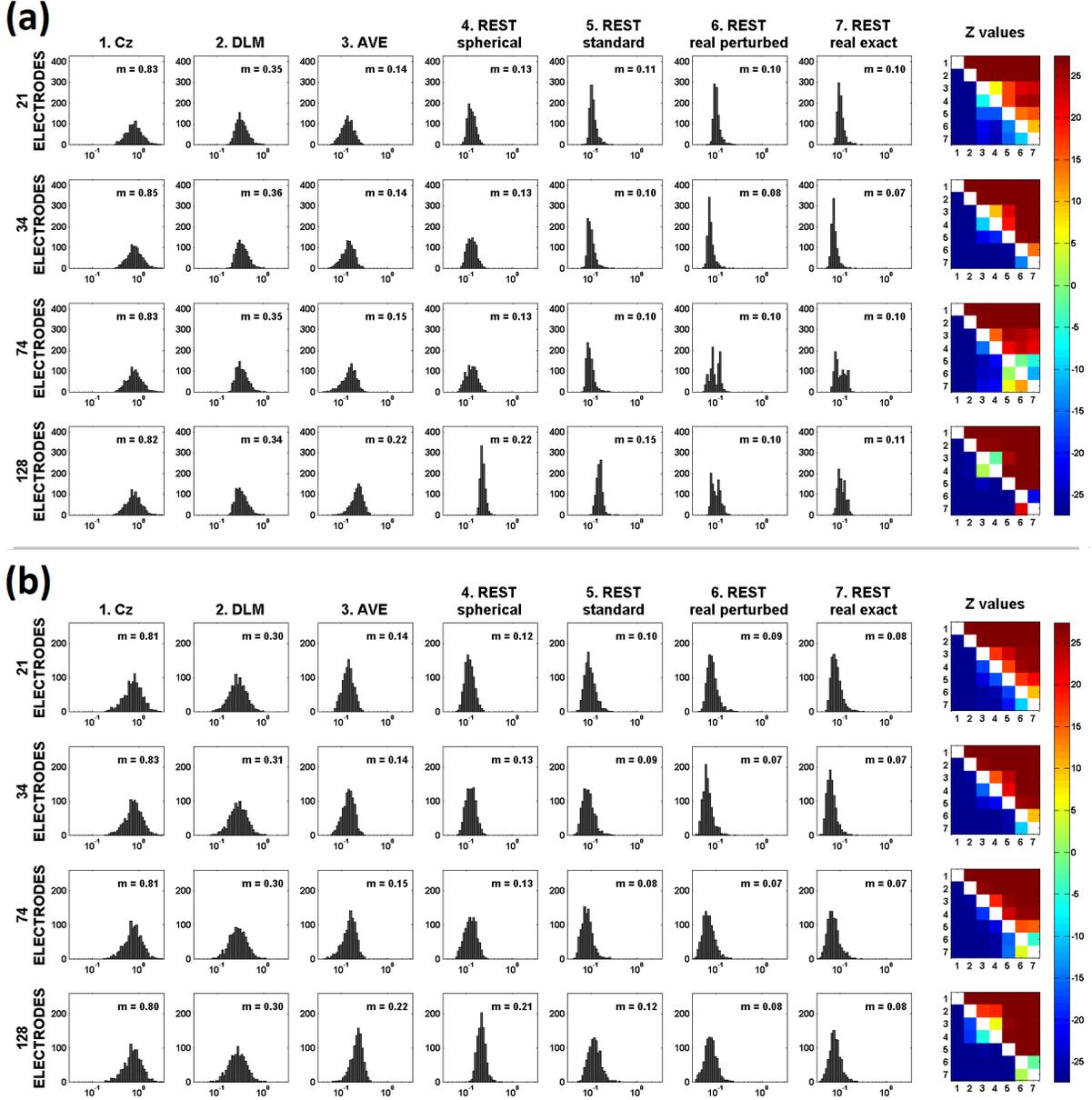}
\caption{\small{Histograms of the relative error ($RE$) for EEG potentials corrupted by either white Gaussian instrumentation noise (panel a) or iso-spectral instrumentation noise (panel b), i.e.~for all combinations of number of EEG electrodes and EEG referencing conditions. The SNR was set to 10. The histograms collect the data from 1000 simulation repetitions. The mean value of $RE$ is denoted by $m$. For the ease of visualization, the abscissa values for the histograms have been scaled logarithmically. Rightmost side of each panel: z-values for non-parametric paired sample statistics, i.e.~Wilcoxon signed-rank test, performed for the contrast of the $RE$ distributions obtained in different EEG referencing conditions, here labelled with a progressive number from 1 to 7, and for a specific electrode density.}}
\label{noisy_pot_hist}
\end{figure*}

The above scenario changes when we turn to the case of EEG potentials corrupted by either white Gaussian or iso-spectral instrumentation noise (SNR=10). The corresponding results are summarized in figure \ref{noisy_pot_hist}. For the case of white Gaussian noise (panel a), we first observe an overall increase of the $RE$ for all of the investigated reference conditions, except for the AVE reference, whose performance does not substantially change due to the addition of noise, especially for denser electrode arrays. In the comparison between the different reference performances, REST still remains the best choice, although it must be noted that the $RE$ for the REST reference becomes similar to the one of the AVE reference if a spherical head model is assumed. The observed differences in the RE distributions are all significant at the $p<0.001$ level (Wilcoxon signed-rank test), or at the $p<0.05$ level only for the contrast between AVE and REST \textit{spherical} when 128 EEG electrodes are used, except for the contrast between REST \textit{standard} and REST \textit{real perturbed} when 74 EEG electrodes are used (not significant).\\
When we come to the case of iso-spectral noise (panel b), we observe small changes in the performances of the various EEG references due to the modified noise conditions. Specifically, there is an overall slight decrease in the mean value of $RE$ in comparison to the case of white Gaussian noise, except for the AVE reference, for which we already noted negligible effects due to the addition of simulated instrumentation noise. These small changes, however, do not affect the contrast between the different EEG reference performances. Indeed, REST still remains the best choice of reference, or at least it is comparable to the AVE reference if a spherical head model is assumed. The observed differences in the $RE$ distributions are all significant at the $p<0.001$ level, or at the $p<0.05$ level only for the contrast between REST \textit{real perturbed} and REST \textit{real exact} when 128 EEG electrodes are used.

\subsubsection{Coherency maps.}
\label{sec:res_coh}
We examined the $RE$ distributions for the imaginary part of coherency maps at 10 Hz (which we recall to be the main frequency of the simulated source signals) estimated from the simulated EEG datasets. While differences were found for potentials in the contrast between the noiseless and the noisy case, no noteworthy differences emerged for the imaginary part of coherency maps. This is conceivably due to the fact that the contribution to coherency of the simulated noise, being either white Gaussian or iso-spectral noise, rapidly approaches zero as the average over signal segments of equation (\ref{eq:cohy1}) is performed. Indeed, for the sake of completeness, in the following we will discuss the more general case of the imaginary part of coherency maps derived from noisy potentials.

\begin{figure*}[!p]
\centering
\includegraphics[width=16cm]{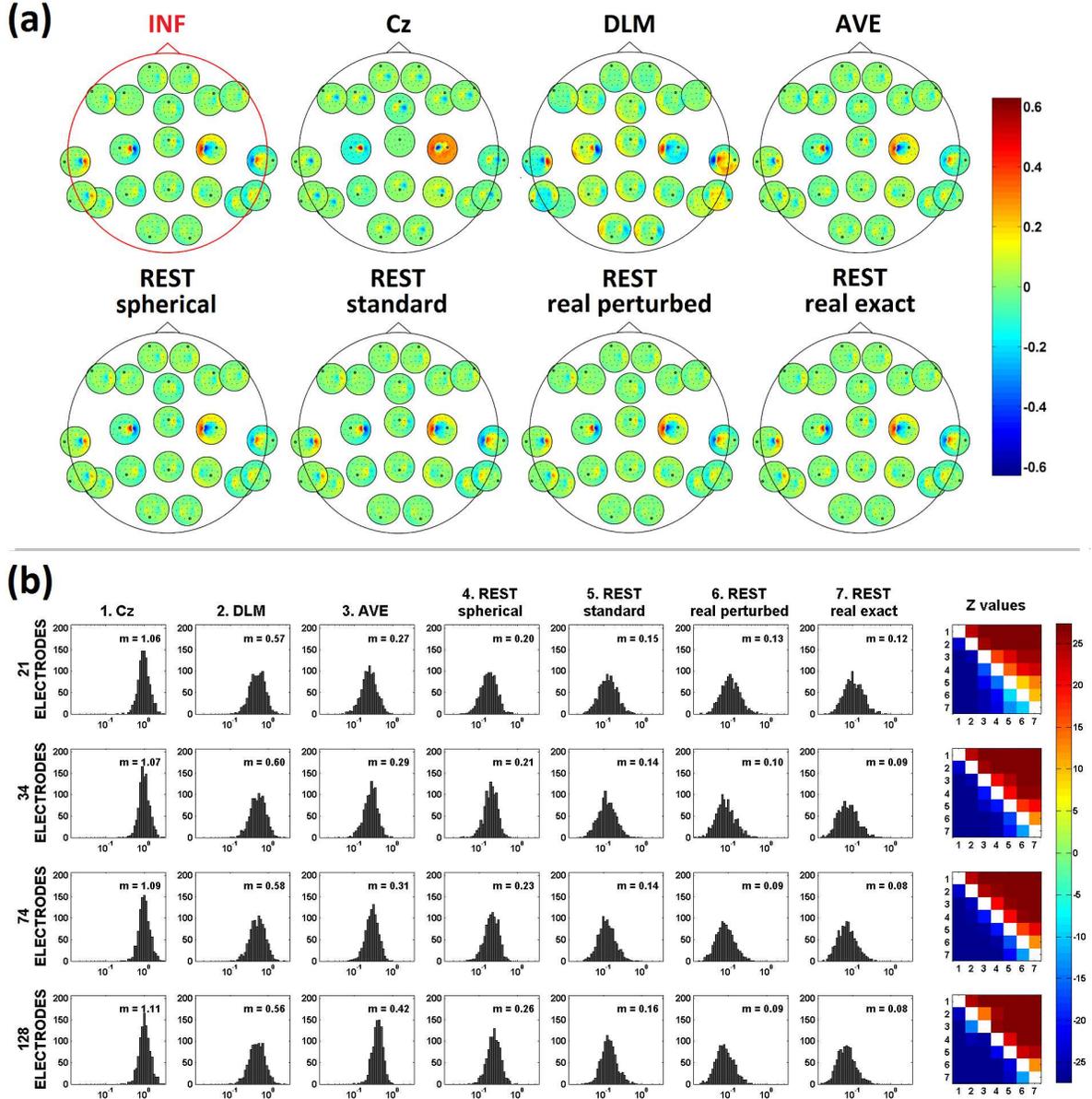} 
\caption{\small{Panel (A): imaginary part of coherency maps at 10 Hz for one example of simulation repetition. These maps have been shown only for the full 128 electrode set. The map of the imaginary part of coherency for the theoretical reference at infinity (INF) is shown, and can be compared to the ones estimated for Cz, DLM, AVE, REST \textit{spherical}, REST \textit{standard}, REST \textit{real perturbed} and REST \textit{real exact} reference. Panel (b): histograms of the relative error ($RE$) for the imaginary part of coherency estimated from the EEG recordings with additive white Gaussian instrumentation noise (SNR=10). The histograms are shown for all combinations of number of EEG electrodes and EEG referencing conditions, and collect the data from 1000 simulation repetitions. The mean value for $RE$ is denoted by $m$. For the ease of visualization, the abscissa values for the histograms have been scaled logarithmically. In the rightmost side of the panel (b): z-values for non-parametric paired sample statistics, i.e Wilcoxon signed-rank test, performed for the contrast of the $RE$ distributions obtained in different EEG referencing conditions, here labelled with a progressive number from 1 to 7, and for a specific electrode density.}}
\label{imcoh_repr_hist} 
\end{figure*}

To illustrate the effects of the reference choice on the imaginary part of coherency, we first discuss an exemplary case chosen from all the simulation repetitions. Specifically, we consider the case in which the EEG recordings were generated by two interacting source dipoles of unit strength, vertically oriented and located in proximity of the left and right supramarginal gyri. In this example, the EEG sensor recordings were corrupted by white Gaussian noise. Figure \ref{imcoh_repr_hist}a shows the corresponding maps of the imaginary part of coherency at 10 Hz, derived from $\mathbf{V}_{INF}$ (INF) (see section \ref{sec:gendata}) and for all the simulated references.
For the visualization of the connections between all pairs of electrodes, we used the following procedure originally introduced by \citeasnoun{nolte04}. The single large circle is the two-dimensional representation of the scalp. At the location of some electrodes (i.e.~merely chosen among all the electrodes for the sake of visualization) small circles are placed representing the scalp and containing the imaginary part of coherency of the respective electrode (marked as a black dot) with all other 128 electrodes. These maps have been here shown only for the full 128 electrode set, whereas the maps for the other electrode subsets, i.e.~including 21, 34 and 74 electrodes, have not been shown. From a qualitative comparison of these maps, we observe that the imaginary part of coherency for all the REST referencing conditions show a spatial pattern which is very similar to that for INF. Small differences can be found for the AVE reference, while the major differences exist for the DLM and Cz reference.

The above observations are supported by a quantitative comparison of the distributions of the $RE$ values from all the 1000 simulations repetitions. In figure \ref{imcoh_repr_hist}(b), we show the histograms of the $RE$ for the imaginary part of coherency estimated from the EEG recordings with additive white Gaussian noise. Similarly to what we have done in figures \ref{noiseless_pot_hist} and \ref{noisy_pot_hist} for potentials, the histograms of the $RE$ are shown for all the combinations of electrode numbers and EEG referencing conditions. Overall, we observe that $RE$ for the imaginary part of coherency has the same basic features which were discussed in the previous section in relation to noiseless potentials. In particular, we are interested in the contrast of different EEG references that can be directly inferred from the z-score values (all significant at the $p<0.001$ level) from pairwise comparisons of the $RE$ distributions, which are shown in the rightmost side of figure \ref{imcoh_repr_hist}(b). We can observe that the largest $RE$ is obtained when Cz or DLM are used as a reference. Lower values were achieved for the AVE reference, although the REST provides superior performances than all the other EEG references in reducing the bias of the reference choice on the estimation of the imaginary part of coherency.

Similar results can be observed when we turn to the case of iso-spectral noise corrupting signals. The corresponding histograms of the $RE$ for the imaginary part of coherency have been shown in figure \ref{isonoise_imcoh} of the Supplementary Material. In the contrast between the two simulated noise conditions, no substantial differences were found in the $RE$ distributions for the Cz, DLM and AVE references. We observed a slight increase of the $RE$ (on average) for REST for the case of iso-spectral noise, regardless of the assumed head model, although the REST still remains the best choice of reference, in comparison to the other EEG references.


\subsection{Real EEG recordings}
The imaginary part of coherency matrices for spontaneous eyes-open resting state activity were estimated at the individual frequency within the Alpha band where the imaginary part of coherency showed its maximum. The analysis of the effects of the EEG reference choice on connectivity mapping and on the estimation of network properties, i.e.~node degree and local efficiency, derived from graph theoretical analysis on the above matrices are presented below.

\subsubsection{Coherency maps.}
\label{sec:realdatacoh}
\begin{figure*}[!p]
\centering
\includegraphics[width=16cm]{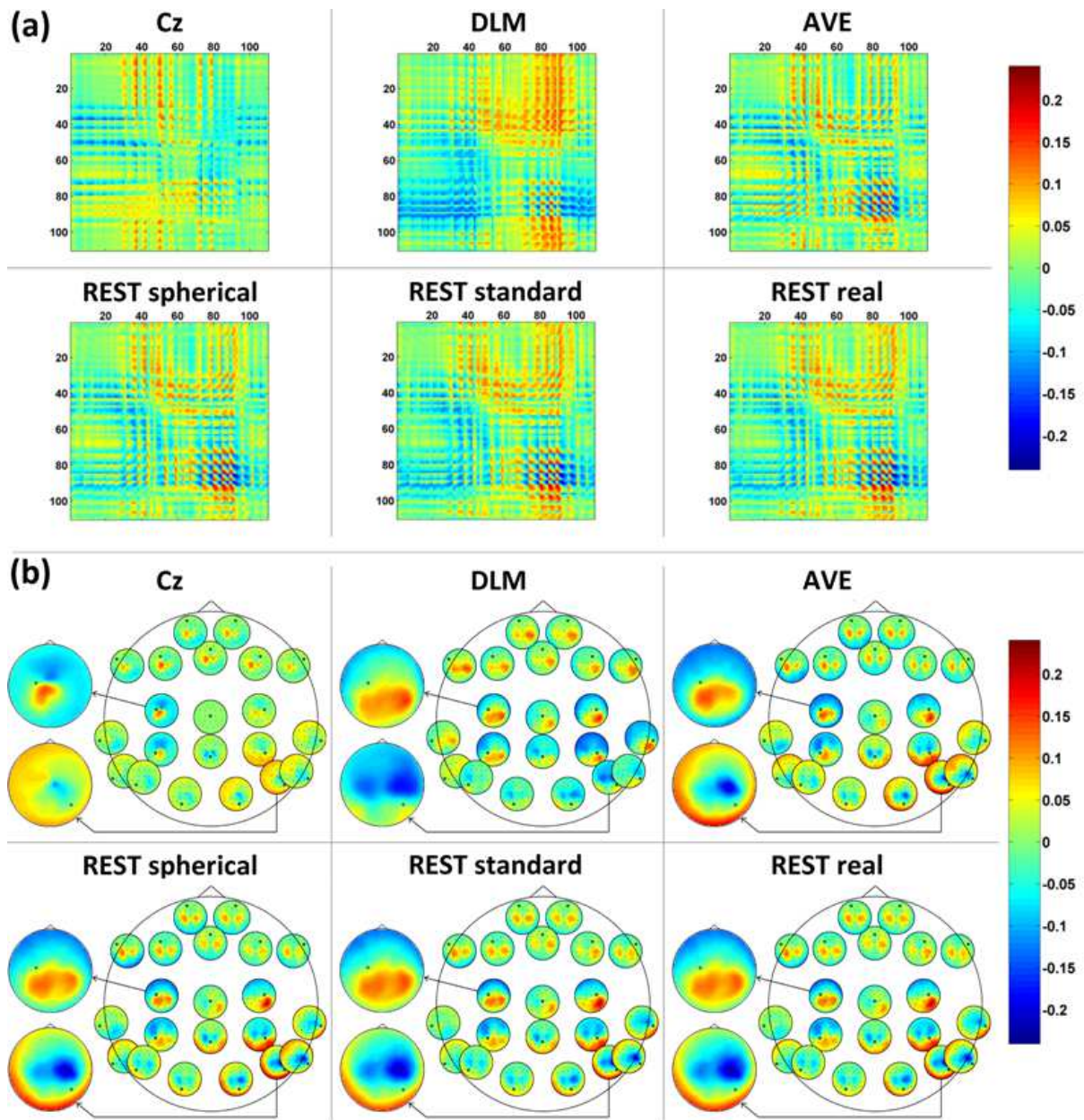}
\caption{\small{Imaginary part of coherency at the individual (per subject) frequency within the Alpha band where the imaginary part of coherency shows its maximum. Data have been averaged over subjects. the group average all-to all-connectomes based on the imaginary part of coherency between all EEG electrodes (panel a) and the corresponding maps (panel b, including a detailed view of the maps for the imaginary part of coherency with respect to channels C3 and P8) are shown for Cz, DLM, AVE, REST \textit{spherical}, for REST \textit{standard} and REST \textit{real} reference.}}
\label{imcoh_realdata}
\end{figure*}

The results of coherency analysis for different EEG referencing conditions are summarized in figure \ref{imcoh_realdata}.
Here, we show the group average all-to-all connectomes based on the imaginary part of coherency between pairwise EEG electrodes (panel a) and the corresponding scalp maps (panel b) for the actual EEG measurements. Unlike simulations, where the theoretical EEG data referenced at infinity were available, for real data it is difficult to assess which EEG reference best reflects the actual brain dynamics. Here, based on simulation results, we argue that REST \textit{real} reference provides the best approximation of the reference at infinity and, thus, can be chosen as golden standard for comparison with other referencing conditions.

In the comparison of the connectomes from different referencing conditions (panel a), we observe macroscopic differences in the connectivity structures revealed by these matrices. In order to quantitatively evaluate these differences, we used a dissimilarity measure defined as one minus the squared Pearson correlation coefficient, i.e.~$d_r=1-r^2$, between the unfolded imaginary part of coherency matrix obtained for REST \textit{real} and those obtained for the other referencing conditions. Based on the above definition, $d_r$ ranges from 0 (no dissimilarity) to 1 (complete dissimilarity). The obtained values for $d_r$ are listed in the following: $d_r$=0.80 for the Cz reference, $d_r$=0.33 for the DLM reference, $d_r$=0.26 for the AVE reference, $d_r$=0.06 for the REST \textit{spherical} reference, and $d_r$=0.01 for the REST \textit{standard} reference. For a comprehensive comparison between all the EEG referencing condition pairs, the values of $d_r$ resulting from the contrast of all pairwise combinations of the EEG references are shown in figure \ref{imcoh_dissimilarities} of the Supplementary Material.

Greater insight into the effect of the EEG reference choice on the estimation of imaginary part of coherency can be obtained by looking at the differences between the connectivity maps which are shown in panel b of figure \ref{imcoh_realdata}. The imaginary part of coherency pattern obtained for REST \textit{real} reference reveals an interesting interaction structure: the central electrodes are mostly interacting with the frontal and occipito-parietal ones, and vice versa. According to our considerations, this pattern has a straightforward interpretation in terms of the underlying brain interaction dynamics, that is, it reveals an interaction occurring between brain sources located in the central regions with other sources located in the frontal and occipital/central regions. Cz reference, in its turn, provides an interaction pattern which looks substantially different from the one obtained for REST \textit{real}, with subsequent possible difficulties in the interpretation of actual brain interaction dynamics. Overall, DLM reference provides better results than Cz reference, even though the imaginary part of coherency maps are slightly shifted to the right, while major differences can be observed in the proximity of the left and right mastoids. AVE references shows a connectivity pattern which is very similar to the one obtained for REST \textit{real}, while no difference can be visually appreciated in the contrast between REST \textit{spherical}, REST \textit{standard} and REST \textit{real}.

\subsubsection{Systematic differences in Node Degree and Local Efficiency.}
The  analysis of the effects induced by the reference on the estimation on network properties, such as node degree and local efficiency, are presented below.

On the main diagonal of figure \ref{degree}, we show, for each of the EEG referencing conditions, the patterns of the average node degree over the 10 subjects.
\begin{figure*}[!htb]
\centering
\includegraphics[width=13cm]{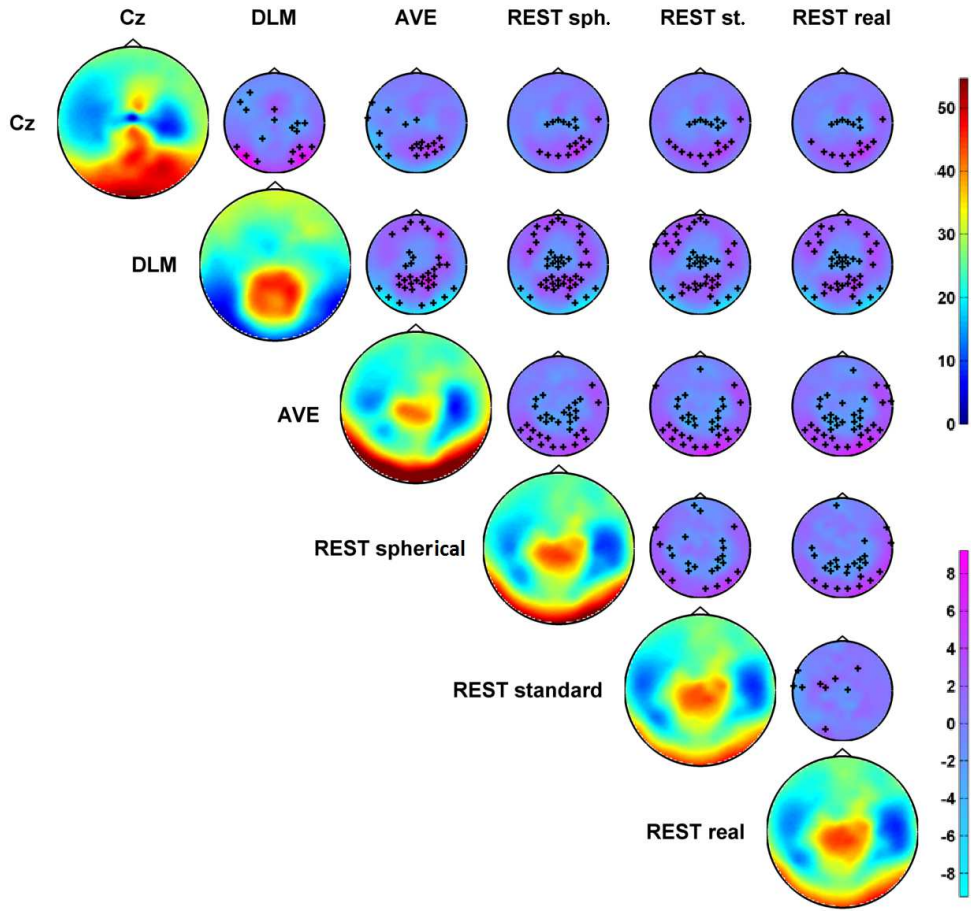}
\caption{\small{Impact of the EEG reference choice on the computation of the Node Degree. Main diagonal: topographical maps of the Node Degree for the different EEG referencing conditions. Off-diagonal: topographical maps of t-values for the contrast between the different referencing conditions by using a paired sample t-test; the black crosses mark the channels showing significant differences at the $p<0.05$ level based on a permutation test (10000 randomizations).}}
\label{degree}
\end{figure*}
As for coherency mapping, systematic differences arise from the contrast between different referencing conditions, which are only due to the choice of that particular reference. Specifically, REST \textit{real} reference reveals a higher degree of connectivity for the electrodes located on the central and occipital regions. A similar pattern can be observed when REST \textit{standard} reference is adopted. On the contrary, a noteworthy increase of connectivity on the occipital electrodes arises when REST \textit{spherical} or AVE references are chosen, whereas more widespread differences can be observed for DLM and Cz references. On the off-diagonals of figure \ref{degree}, we show the maps of t-values for the contrast between the different referencing conditions by using a paired sample t-test. Here, we marked with a cross the channels showing significant differences at the $p<0.05$ level based on a random permutation test. Statistical analysis confirmed the existence of systematic differences in the node degree value among the different referencing condition, with differences revealing specific spatial topographies.  

\begin{figure*}[!htb]
\centering
\includegraphics[width=13cm]{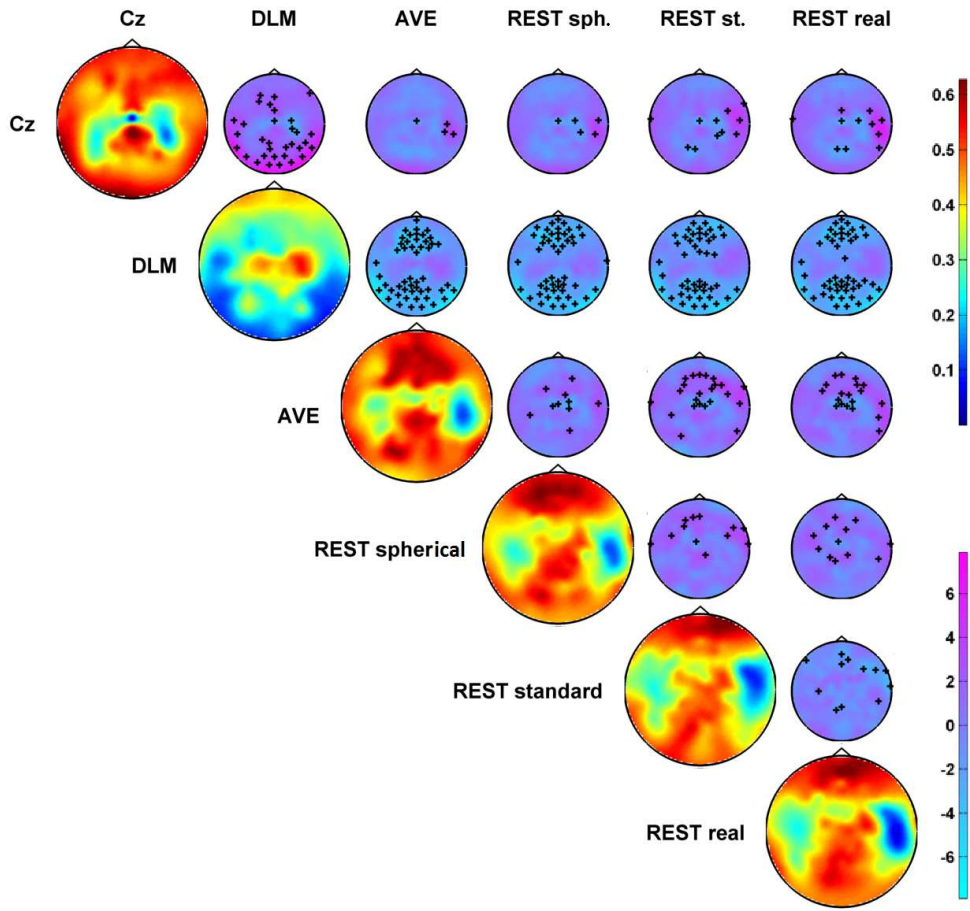}
\caption{\small{Impact of the EEG reference choice on the computation of the Local Efficiency. Main diagonal: topographical maps of the Local Efficiency for the different EEG referencing conditions. Off-diagonal: topographical maps of t-values for the contrast between the different referencing conditions by using a paired sample t-test; the black crosses mark the channels showing significant differences at the $p<0.05$ level based on a permutation test (10000 randomizations).}}
\label{loc_eff}
\end{figure*}

Similar considerations apply to the analysis of the local efficiency. Indeed, also in this case, significant differences can be observed by contrasting the patterns of the local efficiency for different referencing conditions, as illustrated in figure \ref{loc_eff}. These differences are also characterized by specific spatial topographies.

\section{Discussion}

The aim of this study was to investigate the effects of the reference choice on scalp EEG connectivity estimation, including the analysis of the imaginary part of coherency and the characterization of functional network topology based on graph analysis applied to the data. This was first assessed in simulations, where four commonly used reference schemes, i.e.~the  Cz, the DLM, the AVE and the REST reference, were compared to the case of the reference to a point located at infinity, which behaves as the ideal EEG reference \cite{yao01,nunez06}. Specifically, we evaluated the distortion induced in the values of imaginary part of coherency due to the use of the above references, and we examined the effects of the electrode density, sensor noise and, of specific interest for REST, of the head model accuracy. For a direct comparison with previous studies, the effects on EEG potentials were considered as well in simulations.

We found that the Cz reference substantially alters EEG potentials and imaginary part of coherency values in comparison to all the other references. This is essentially due to the influence of the electrical activity at the reference site which, as expected, is non-neutral, thus resulting into a mismatch between the potential and connectivity values referenced to Cz and the respective values referenced to infinity. It is reasonable to expect that similar considerations apply to other cephalic references, e.g.~the nose reference, which have not been explicitly addressed in this study but are potentially affected by the same issue. Although it has been sometimes argued that the mastoids are relatively inactive and, thus, the DLM might be a suitable choice for the reference, this was shown to be false \cite{dien98,hagemann01,nunez06}, and also the findings of the present study do not support this view. Indeed, the DLM reference, while showing a better performance than the Cz reference, induces significantly larger distortions on EEG potential and imaginary part of coherency if compared to the AVE and REST references. We found that the electrode density does not affect the performances of Cz and DLM reference, which is reasonably due to all the electrodes being equally influenced by the reference activity. Moreover, the distortion of EEG potentials is enhanced by additional either white Gaussian or iso-spectral sensor noise. This can be regarded to as the effect of additional noisy activity on the reference signal. 

Although the AVE reference is often acknowledged as a quite neutral reference if used with a large number of electrodes, our findings showed that also the AVE reference is not completely free of biases, mainly due to the due to potential sampling being limited to the upper part of the head. Indeed, we found that the relative error for EEG potentials and imaginary part of coherency increases for increasing sensor density, reasonably due to the scalp coverage being still inadequate. These findings are in line with those of previous studies \cite{dien98,nunez06}. Interestingly, the AVE reference performance is not changed by adding noise, especially for denser electrode arrays. This can be motivated by the fact that the contribution of the simulated noise to the average over signals rapidly decrease as the number of averaging signals increases.

Our simulations demonstrated that the off-line transformation of EEG recordings performed by REST, in the attempt to estimate the scalp potentials with respect to infinity, substantially reduces the above reference effects. Since it has been argued that the REST reference performance might depend on the electrode density and head model uncertainty, we have demonstrated that, for a number of electrodes ranging from 21 to 128 and for various levels of accuracy in the knowledge of the head model, REST successfully reduces the bias introduced by other references. In particular, we showed that the availability of high-density EEG systems and an accurate knowledge of the head model are crucial elements to improve REST performance, in agreement with the findings of previous studies \cite{zhai04,liu15}. In addition, we showed that a realistic head model based on the individual head anatomy is preferable to the one based on a standard head anatomy, especially when high density EEG is available. We also found that REST is sensitive to additional white Gaussian or iso-spectral sensor noise. This is essentially due to REST assuming the sources of the EEG recordings lying inside the equivalent source distribution (ESD). Since the instrumental noise is not generated by sources inside the ESD, the effectiveness of the standardization to a reference point at infinity becomes less accurate in comparison to the noiseless case \cite{zhai04}. However, it must be noted that REST performs better than AVE reference even in presence of noise, except when a non-realistic three-shell spherical head model is used, for which the performances of REST and AVE reference were found to be similar.

In this work, a particular emphasis has been placed on the comparison between REST and AVE reference, the superiority of one method over the other having been argument of some debates \cite{kayser10,nunez10}. Based on the findings of our simulations, we concluded that REST can provide superior performances than AVE reference in reducing the reference bias if a head model based on either a standard or individual head anatomy is assumed, or even if an idealized (three-concentric sphere) head model is assumed and the noise is adequately suppressed.

The analysis performed on real EEG data recorded during eyes-open resting state confirmed that the choice of the reference has a non-negligible effects on EEG connectivity analysis performed at sensor level. Since in actual experiments the EEG potentials referenced to infinity are not available, we evaluated the reference effects in comparison to the REST performed by using a realistic head model based on subject's anatomy. Our findings highlighted a systematic change of the spatial pattern of functional connections estimated between scalp EEG electrodes depending on the chosen reference, consistently with the results from previous studies \cite{marzetti07}. The distortion of connectivity patterns was larger for the Cz reference, and progressively decreases when using, in turn, the DLM, the AVE, the REST \textit{spherical} and the REST \textit{standard} references. Strikingly, we also showed that the network attributes that rely on local graph properties, i.e.~node degree and local efficiency, are significantly influenced by the EEG reference choice. This result extends previous findings on
the dependence of network pattern and weighted density on the chosen reference \cite{qin10}. Overall, the above results raise non-trivial issues for the interpretation of scalp connectivity measures in terms of the underlying brain interaction dynamics. Especially, it must be noted that one should not treat the findings of different reference schemes as interchangeable, inasmuch as the choice of a particular reference induces significant and systematic changes in data analysis results.

\subsection{General comments on reference-free approaches}

Besides the methods concerned in this paper, when dealing with the issue of the EEG reference, the availability of reference-free techniques should also be considered. For instance, the Surface Laplacian (SL) \cite{hjorth75,kayser15,nunez06} is a mathematical transformation applied to the EEG scalp potentials which is not biased by the reference effects. Indeed, the SL relies on the estimation of the spatial second derivatives of scalp EEG potentials, i.e.~through either a nearest-neighbour approach \cite{hjorth75,hjorth80} or a more accurate spline interpolation approach \cite{perrin89}. As a consequence, the SL is not affected by the addition (or subtraction) of a constant value to the potentials measured by all the EEG electrodes, which is, by itself, the act of referencing potentials. Despite of this advantage, however, the SL has the limitation of suppressing the activity of deep and distributed brain sources. This is essentially due to the spatial derivative acting as a high-pass spatial filter, which tends to isolate effects due to shallow and localized sources rather than to deep and distributed sources. Similar arguments apply to the Current Source Density approach \cite{nunez06}, which also relies on the estimation of the second derivatives of scalp potentials, and thus has the inherent limitation of suppressing broad scalp activities, which are actually very common in EEG.
 
A different approach consists in the so-called bipolar EEG recordings. This approach is more popular in clinical work than in cognitive studies, and is routinely employed in the interpretation of scalp as well as intracranial EEG \cite{niedermeyer05,zaveri06}. Bipolar recordings consist in the measurement of the potential difference between pairs of closely spaced electrodes. The more the electrodes of any pair are close to each other, the better the recorded potential difference approximates the local gradient of the electric potential in the direction between the electrodes, which is roughly proportional to the current density tangential to the scalp \cite{srinivasan96}. In conventional bipolar schemes (or montages), e.g.~the ``double banana'', the electrode pairs are chosen in a sequential manner, i.e.~the second electrode of the first pair is also the first electrode in the next pair (e.g., Fp1-F3, F3-C3, C3-P3, P3-O1, and so on). This strategy implicitly overcomes the issue of the EEG reference, since the contribution of the reference electrode is removed when computing the difference between the potentials of any pair of electrodes. In contrast to referential recordings, on which we specifically focused in this study, bipolar recordings can be a more effective strategy to remove artifact contamination, identify local activations, and provide a reference-free representation of phenomena under observation \cite{zaveri06}. However, similarly to the SL, this results in a suppression of large and distributed activations, due to the effect of spatial derivatives which is equivalent to a high pass spatial filter.

Another strategy to get rid of the reference effects is to perform the connectivity analysis at the source level. Indeed, it has been shown that the choice of the EEG reference does not affect the inverse localization of neural active sources, at least for noiseless potentials \cite{geselowitz98,pascualmarqui93}. Thus, once provided a solution to the EEG inverse problem, connectivity can be directly estimated between the activities of localized brain sources. This approach, however, raises the question of how accurate is the brain source reconstruction. It is well known, for instance, that high density EEG should be preferred over low density EEG to perform a reliable source reconstruction. The advantages and limitation of this approach will not be addressed here, as they go beyond the scope of this paper. Our aim was here to highlight how the choice of the reference affects the estimation brain connectivity inferred from scalp EEG, which still remains a standard practice for many research or clinical applications \citeaffixed{carlino15,herrera-diaz15,vanstraaten15,ligeza16,naro16,wang16,yuvaraj16}{e.g.,}.


\section{Conclusions}
In conclusion, the results of the present study have demonstrated that different references significantly alter the topography of EEG connectivity patterns. Accordingly, the choice of the EEG reference introduces a bias on the interpretation of these patterns in terms of brain interactions, as well as the characterization of network topology which can be derived from graph theoretical analysis applied to these data. These findings, which have been obtained by analysing the imaginary part of coherency estimated from the whole signal length, can be generalized to other connectivity metrics relying on either temporal or spectral properties of the data. This includes the study dynamic functional connectivity, i.e.~functional connectivity varying as a function of time. In this case, we expect significant changes in the connectivity measures for each time interval in which the connectivity is observed, with subsequent difficulties in interpreting the results in terms of the time-varying properties of brain interactions.\\
In order to reduce the effects of the reference choice on the analysis of EEG connectivity, we recommend the use of the REST reference. This approach will not only allow for an unbiased (or at least a less biased) analysis of the EEG data, but also facilitate the comparison of results obtained from different laboratories or stored with different references in databases collected over time, which is of fundamental importance for cross-laboratory studies and in clinical practice.

\section*{Acknowledgements}
The authors would like to acknowledge Prof. Guido Nolte (University Medical Center Hamburg-Eppendorf, Hamburg, Germany) for sharing part of the code used in this study, and Prof. Gian Luca Romani for helpful comments. This work was supported by the Italian Ministry of Education, University and Research (PRIN 2010-2011 n.~2010SH7H3F\_006 ``Functional connectivity and neuroplasticity in physiological and pathological aging''), by the Italian Ministry of Health (GR-2011-02351822) and by the European Commission (grant ``BREAKBEN - Breaking the Nonuniqueness Barrier in Electromagnetic Neuroimaging'', H2020-FETOPEN-2014-2015/H2020-FETOPEN-2014-2015-RIA, Project reference: 686865). 

\subsubsection*{Conflict of interest.}
The authors disclose any actual or potential conflict of interest.

\section*{References}
\bibliography{references}

\end{document}